\documentclass[10pt, journal, final]{IEEEtran}
\usepackage{amsmath,graphicx,multirow,pifont}
\usepackage{microtype}

\def\p{{\mathbf{p}}}
\def\x{{\mathbf{x}}}
\def\w{{\mathbf{w}}}
\def\D{{\mathcal{D}}}
\def\A{{\mathcal{A}}}
\def\V{{\mathcal{V}}}
\newcommand{\xmark}{\ding{55}}
\newcommand{\cmark}{\ding{51}}

\title{Modular End-to-end Automatic Speech Recognition Framework for Acoustic-to-word Model}
\author{Qi Liu, \IEEEmembership{Student Member, IEEE}, Zhehuai Chen, Hao Li, Mingkun Huang, \\ Yizhou Lu, and Kai Yu, \IEEEmembership{Senior Member, IEEE}
\thanks{
This work has been supported by National Key Research and Development Program of China (Grant No.2017YFB1002102). The authors would like to thank Heinrich Dinkel and Rao Ma for English proofreading and editing. (\emph{Corresponding author}: Kai Yu.)

Qi Liu, Zhehuai Chen, Hao Li, Mingkun Huang, Yizhou Lu, and Kai Yu are with the SpeechLab, Department of Computer Science and Engineering, and MoE Key Lab of Artificial Intelligence, AI Institute, Shanghai Jiao Tong University, Shanghai, China. (e-mail: liuq901@sjtu.edu.cn; chenzhehuai@sjtu.edu.cn; lh575526@sjtu.edu.cn; mingkunhuang@sjtu.edu.cn; luyizhou4@sjtu.edu.cn; kai.yu@sjtu.edu.cn)
}}

\begin{document}

\maketitle

\begin{abstract}
End-to-end (E2E) systems have played a more and more important role in automatic speech recognition (ASR) and achieved great performance. However, E2E systems recognize output word sequences directly with the input acoustic feature, which can only be trained on limited acoustic data. The extra text data is widely used to improve the results of traditional artificial neural network-hidden Markov model (ANN-HMM) hybrid systems. The involving of extra text data to standard E2E ASR systems may break the E2E property during decoding. In this paper, a novel modular E2E ASR system is proposed. The modular E2E ASR system consists of two parts: an acoustic-to-phoneme (A2P) model and a phoneme-to-word (P2W) model. The A2P model is trained on acoustic data, while extra data including large scale text data can be used to train the P2W model. This additional data enables the modular E2E ASR system to model not only the acoustic part but also the language part. During the decoding phase, the two models will be integrated and act as a standard acoustic-to-word (A2W) model. In other words, the proposed modular E2E ASR system can be easily trained with extra text data and decoded in the same way as a standard E2E ASR system. Experimental results on the Switchboard corpus show that the modular E2E model achieves better word error rate (WER) than standard A2W models.
\end{abstract}


\section{Introduction}
Deep learning has been widely used in ASR systems. Traditional ASR with deep learning always employs ANN-HMM hybrid systems \cite{morgan1990,bourlard1992,morgan1995}: ANN predicts the posteriors of HMM states and HMM is trained separately to fit the long term model. Hybrid systems have some disadvantages. First, forced alignment is needed to train the ANN part in the traditional ANN-HMM pipeline. Second, the short-term model ANN and long-term model HMM are trained separately. Therefore, the knowledge that they have learned will not be shared with each other. Finally, a decoding mechanism such as the weighted finite-state transducer (WFST) \cite{mohria2002} and lattice generation/search is needed during the test phase. However, with more powerful networks like long short-term memory (LSTM) \cite{sak2014} and convolutional neural network (CNN) \cite{qian2016}, the E2E systems have the ability to directly model the word sequence from the acoustic feature with only neural network computing.

Connectionist temporal classification (CTC) \cite{graves2006} is a widely used E2E algorithm. CTC adds a special label \textbf{blank} to model the intermediate frame. The \textbf{blank} label and forward-backward algorithm enable CTC to convert unsegmented input sequences to varied-length output sequences. CTC based systems have achieved good results in several sequence labeling tasks including speech recognition \cite{graves2014} and handwriting recognition \cite{liu2015}.

Sequence to sequence (S2S) \cite{sutskever2014} is another type of E2E model, which consists of two networks: an encoder and a decoder. The encoder models the input sequence as embedding vectors and the decoder uses these vectors to generate the output sequence. In recent years, S2S models, especially attention-based S2S models, have achieved great success on both natural language processing \cite{cui2017} and speech recognition \cite{chan2016,bahdanau2016}.

RNN-transducer \cite{graves2012} combines the advantages of both CTC and S2S models. It contains an encoder-decoder like mechanism and a CTC like criterion. Researches have shown that RNN-transducer could achieve competitive word error rates on speech recognition \cite{graves2013,battenberg2017}.

Many researchers focus on E2E systems during the training phase. However, some works such as \cite{liu2015,miao2015} still involve extra decoding mechanisms like WFST. In this paper, we mainly focus on E2E systems during the decoding phase, i.e., the whole system performs like one neural network during test. We called it A2W property or E2E property.

Extra text data can be used to train a strong language model and generate a WFST. It has been shown that the extra text data involved by WFST with language model can significantly improve the performance of hybrid ASR systems. However, E2E systems decode the word sequence directly from acoustic features with only neural networks. This property gives the E2E systems faster decoding speed \cite{chen2016} and the ability to store only neural network parameters, making it possible to deploy on low-resources machines. However, combining WFST and E2E ASR systems brutally will break this property. Therefore many researchers work on how to add large scale text data into E2E ASR systems. \cite{hori2018} joins CTC, attention-based decoder and RNN language model together as a big joint decoder. \cite{hayashi2018} applies back-translation to convert text data to unsupervised acoustic data. \cite{renduchintala2018} uses the phoneme sequence to train a multi-modal E2E system.

Currently, E2E systems based on phoneme, character, or sub-word generated by byte pair encoding (BPE) \cite{sennrich2016} have achieved great performance. However, the desired output of ASR systems is word sequence rather than sequences consisting of these small units. Thus an additional procedure is still needed to combine these small units to words. Moreover, some Asian languages including Chinese, Japanese, and Korean, are more difficult to split into these small units compared with Latin based languages. Therefore, in this paper, we try to investigate how to use extra text data to improve the performance of `true' E2E models, i.e., A2W models.

In this paper, we use a novel modular E2E system \cite{chen2018}. The modular E2E system contains two networks, one is an acoustic-to-phoneme (A2P) network, and the other is the phoneme-to-word (P2W) network. During the training phase, the acoustic data will be used to train both A2P and P2W networks and extra text data can be used to train the P2W network. Finally, during the decoding phase, the output of the A2P network will be the input of the P2W network, which means the proposed model outputs word sequence from input acoustic feature sequence with only neural network calculations. This makes the whole modular system performs like a normal E2E system. Compared with traditional E2E systems, the proposed modular E2E system has three advantages. First, it performs like an A2W model during the decoding phase, without the need to store other parameters such as WFST to decode the results. Second, the modular E2E system can be trained with extra text data while holding the E2E property. The most commonly used method to incorporate text data in A2W models is WFST \cite{miao2015}, which violates the E2E property. Finally, the modular design makes the system easy to extend or adapt. In this paper, we give an example of out of vocabulary (OOV) words extension.

The rest of paper is organized as follows. Section \ref{sec:back} briefly introduces the background of traditional E2E ASR systems. Section \ref{sec:mod-e2e} proposes our modular E2E model. Section \ref{sec:imp} shows the implementation details of modular E2E systems. Finally section \ref{sec:exp} demonstrates the experimental results and section \ref{sec:conc} gives the conclusion.

\section{Traditional E2E ASR Systems}
\label{sec:back}

\subsection{Connectionist Temporal Classification}

E2E system is designed to solve the sequence labeling problem, which predicts a corresponding output sequence from a given input sequence. The difficulty of solving sequence labeling problem with deep learning lies in the mismatch between the lengths of input and output sequences. In practice, an input acoustic feature may contain hundreds of frames, while the corresponding word sequence only has about twenty words.

CTC \cite{graves2006} uses a special label \textbf{blank} to fill in the intermediate frames. Formally, the merge function is defined as 
\begin{equation}
    \beta:[V\cup\{\textbf{blank}\}]^*\rightarrow V^*,
\end{equation}where $V$ is the vocabulary of output word sequence. The merge function will first combine the same consecutive words together and then remove all the \textbf{blank} symbols. For example, $\beta(a--aab)=\beta(--a-ab)=\beta(a-abbb)=\beta(aa-aab)=aab$ where $-$ means \textbf{blank}. In other words, the previous example shows the valid CTC alignment of the word sequence `aab' with an input feature of length 6.

Let $\x$ be the input feature sequence and $\w$ be the corresponding word sequence. The CTC criterion, i.e., the probability $P(\w|\x)$ is the summation of all the possible CTC alignments by using merge function $\beta$: 
\begin{equation}
    P(\w|\x)=\sum_\pi P(\pi|\x)=\sum_\pi\prod_{t=1}^T P(\pi_t|\x),
\end{equation}
here $\pi\in\beta^{-1}(\w)$ and $T=\text{length}(\pi)=\text{length}(\x)$. Since $\pi$ and $\x$ have the same length, $P(\pi_t|\x)$ can be easily calculated by a neural network with a softmax output layer.

\begin{figure}
\centering
\includegraphics[width=\linewidth]{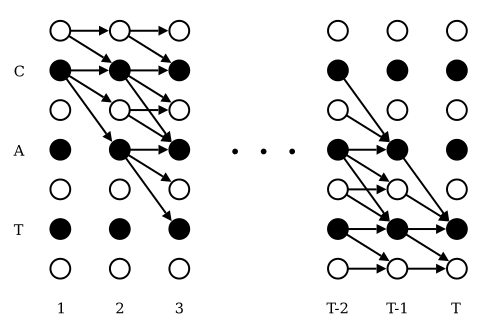}
\caption{The possible paths of forward-backward algorithm \cite{graves2006}. The white circles denote optional \textbf{blank}.}
\label{fig:ctc}
\end{figure}

\begin{figure}
\centering
\includegraphics[width=\linewidth]{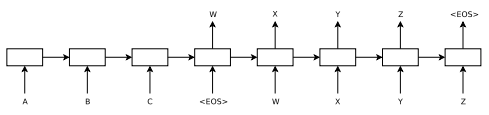}
\caption{An example that the S2S model predicts `WXYZ' with input `ABC' \cite{sutskever2014}. The left part is the encoder network and right part is the decoder network.}
\label{fig:s2s}
\end{figure}

The number of $\pi$ grows exponentially with the input sequence length. Thus $P(\w|\x)$ cannot be calculated efficiently. However, the probability can be calculated by the forward-backward algorithm \cite{baum1967,baum1968}. Figure \ref{fig:ctc} gives a brief overview of applying the algorithm with the CTC criterion, and the details can be found in \cite{graves2006}.

\subsection{Sequence to Sequence}
S2S \cite{sutskever2014}, also known as encoder-decoder, is another commonly used E2E method. CTC predicts a single word or \textbf{blank} for each frame of input feature, and then uses merge function $\beta$ to derive the output word sequence. However, S2S deals with the sequence labeling problem as a conditional language model problem. In other words, S2S is trained as a language model by generating output word sequences conditioned on input features.

Formally, the S2S model contains two networks: encoder and decoder. The encoder `encodes' the input feature to a single compressed embedding vector. The decoder will use the embedding vector and `decode' the output word sequence as a conditional language model.

Let $\x$ and $\w$ be the input feature and output word sequence. The criterion of S2S, i.e. the probability $P(\w|\x)$ is the product of the conditional probability of each single word by chain rule: \begin{equation}P(\w|\x)=\prod_{i=1}^NP(w_i|\x,\w_{1:i-1})\end{equation} here $N=\text{length}(\w)$. The conditional probability can be calculated by two neural networks:
\begin{align}
\textbf{h}&=\text{encoder}(\x) \\
P(w_i|\x,\w_{1:i-1})&=\text{decoder}(\textbf{h}, \w_{1:i-1}).
\end{align} Figure \ref{fig:s2s} illustrates the framework of the S2S model.

However, since the S2S model compresses the input feature to a single embedding vector. When the length of the input sequence is too long, some information, especially earlier information of the input feature might be lost, resulting in performance degradation. Attention mechanism \cite{bahdanau2015} is proposed to utilize the information of input features more efficiently. \cite{chan2016,variani2017} show that attention mechanism can obtain a large improvement on the final WER result. Therefore attention based S2S model is used in our work.

\subsection{E2E ASR Systems}
Currently, two types of E2E ASR systems are commonly used. The first type maps the acoustic feature to a sequence of small units such as phonemes, characters, or bigger sub-word units such as BPE \cite{miao2015,lu2016,miao2016,watanabe2018,rao2017,toshniwal2017}. This type of E2E ASR system has relative small `vocabulary' size (usually less than 100), which enables efficient training with lower memory consumption. In practice, these E2E systems are able to achieve good performance. However, the models of this type usually need an additional module to convert the small unit sequence to the corresponding word sequence. The additional module including WFST \cite{miao2015} or language model rescoring based beam search \cite{watanabe2018} may break the E2E property during the decoding phase. Moreover, a language model trained with extra text data is usually used to build a WFST separately with the neural network. This additional language model, especially WFST, might consume lots of resources that makes the whole system hard to deploy on low resource devices.

This paper mainly focuses on the other type, which is usually called A2W models \cite{audhkhasi2017,audhkhasi2018,yu2018}. It directly maps the acoustic feature to a word sequence. This type can decode the output word sequence with a single neural network. However, due to the large number of output units (usually more than ten thousand words) and long acoustic feature length (usually about one thousand frames), systems of this type are harder and slower to train. Moreover, the large memory consumption makes some algorithms like RNN-transducer hard to implement in a proper A2W system. Another problem is that A2W networks can only be trained with acoustic data. The language model trained with extra text data cannot be easily integrated with A2W models.

Choosing which type of E2E ASR system is a trade-off. The decoding results of A2W models could be generated directly by the neural networks. This indicates that low resource terminals like smartphones or automobiles only need to store the neural network parameters. Many smartphones also have a neural processing unit (NPU) in their system on a chip (SoC), which can speed up the calculation of neural networks. The other type can employ greedy search decoding and word boundaries to achieve the A2W purpose \cite{he2019,li2019}. However, it will abandon the simplicity of using extra text data to improve model performance as the acoustic data is harder to obtain than the text data. Moreover, the proposed modular E2E ASR system is designed to be optimized by both acoustic data and text data while the whole system performs like a single A2W neural network during the decoding phase like \cite{renduchintala2018,sriram2017}.

\section{Modular E2E ASR System}
\label{sec:mod-e2e}
\begin{figure}
\centering
\includegraphics[width=\linewidth]{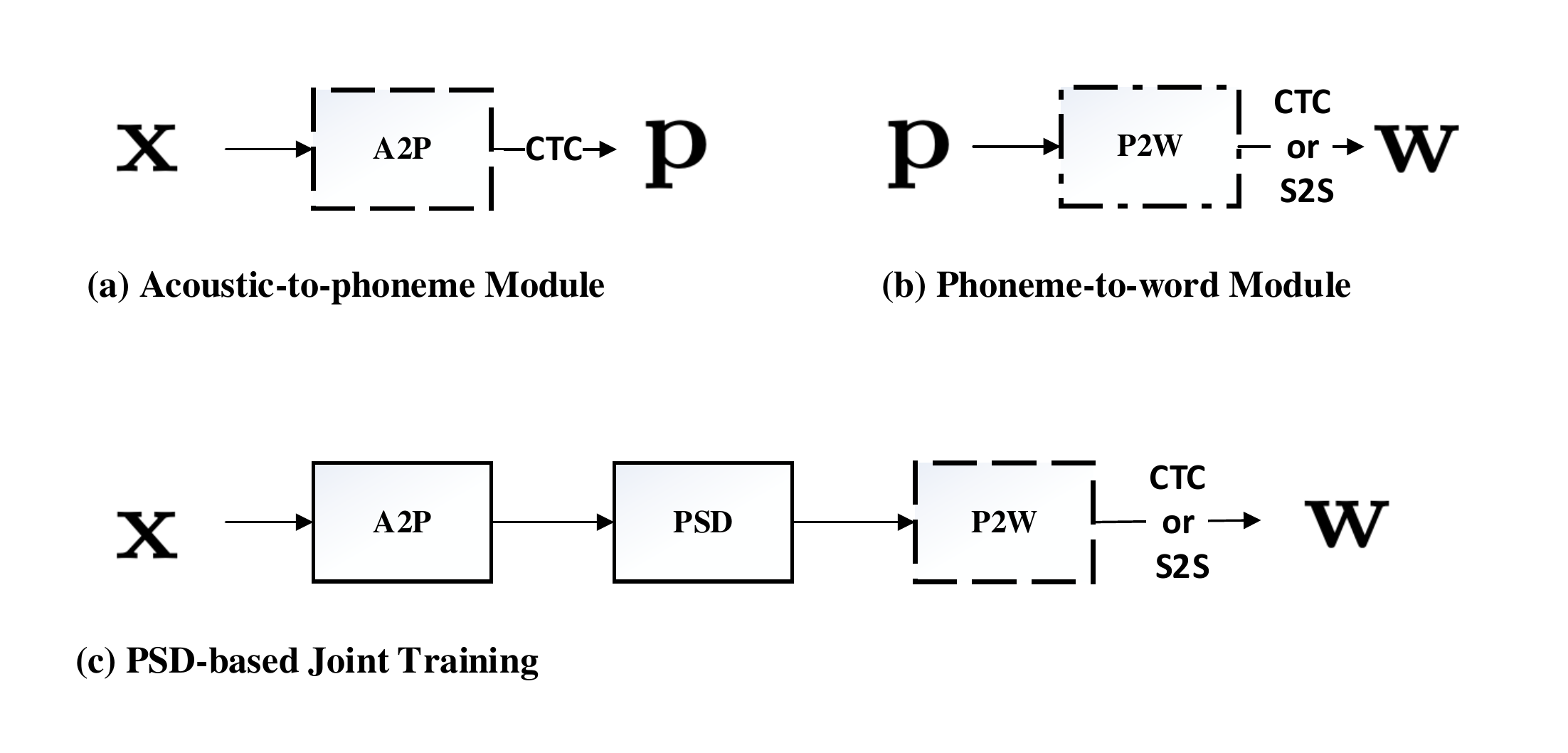}
\caption{The framework of modular E2E systems \cite{chen2018}. The dashed line square indicates the trained part.}
\label{fig:framework}
\end{figure}

In this section, the proposed modular E2E ASR system \cite{chen2018} will be introduced. The modular E2E system consists of two networks: an A2P network and a P2W network. The A2P network can only be trained with acoustic data, such that it predicts the corresponding phoneme sequence with given acoustic features. Meanwhile, the P2W network translates the phoneme sequence to the desired word sequence, which can be trained by both acoustic data and text data. Finally, the P2W network is fine-tuned by the acoustic data. Figure \ref{fig:framework} shows the whole framework of modular systems.

\subsection{A2P Network}
The A2P network is trained on acoustic data, which predicts the posterior probability of phoneme sequence $\p$ with the given acoustic feature sequence $\x$, i.e. \begin{equation}P(\p|\x)=\text{A2P}(\x).\end{equation}

In the modular system, the A2P network can be considered as the acoustic model in the standard ANN-HMM hybrid setting. It recognizes the acoustic features and produces the corresponding phoneme sequence. In fact, it is actually a phoneme-based E2E ASR network, i.e., the first type of E2E ASR system mentioned in the previous section. This indicates that all the E2E optimizing methods can be used to improve the performance of the A2P network. 

Even though all E2E criteria can be used to train the A2P network, in this work, only the vanilla CTC is used as the criterion of the A2P network. It is because that the predicted phoneme sequence might contain many errors. Then the whole posterior sequence will be the output of the A2P network. However, the CTC criterion can predict the posterior sequences only based on the input acoustic data. The S2S criterion must predict the posterior sequences not only relying on the input acoustic data, but also the previously predicted phoneme sequence. More precisely, for a given acoustic feature $\x$ , phoneme sequence $\p=(p_1, \ldots, p_T)$ and a trained A2P network, the CTC criterion is formulated as, \begin{equation}P(\p|\x)=\prod_{i=1}^T\text{A2P}(p_i|\x).\end{equation}Therefore, for any possible $\p$, the posterior $P(\p|\x)$ could be calculated only depending on $\x$. However, for cross entropy criterion used in S2S, we have \begin{equation}P(\p|\x)=\prod_{i=1}^T\text{A2P}(p_i|\x,\p_{1:i-1}).\end{equation}Here, for any possible $\p$, $P(\p|\x)$ depends not only on $\x$ but also $\p$ itself. To provide more information, the output of the A2P network, i.e., the input of the P2W network is the posterior of all possible phoneme sequences rather than a single best phoneme sequence. More precisely, let $V_p$ denote the phoneme vocabulary, the output of the A2P network is a $T\times|V_p|$ matrix which represents the posterior sequence rather than a single length $T$ phoneme sequence. However, as mentioned above, for a given acoustic feature $\x$, S2S models can not produce the $T\times|V_p|$ matrix since it calculates the posterior which not only relies on $\x$ but also on $\p$. Therefore, only CTC is used as the criterion for the A2P network.

\subsection{P2W Network}
The P2W network can be trained by both acoustic data and text data. The input of the P2W network is the posterior of all possible phoneme sequence $\p$ and the output is the desired word sequence $\w$, i.e. \begin{equation}P(\w|\p)=\text{P2W}(\p).\end{equation}
The P2W network can be considered as the language model part of the ANN-HMM hybrid system, although the generated word sequence depends on the input phoneme posterior. Compared with the traditional language model $P(\w)$, the P2W network is trained with $P(\w|\p)$. P2W network and language model are both trained on large scale text data. However, the language model focuses on the unconditional internal relationship among all the words in sentences. Nevertheless, P2W models learn not only the unconditional word distribution but also the phoneme to word dictionary. Additionally, the phoneme alignment, i.e., the alignment about which phoneme belongs to which word, is also learned by P2W models. In other words, the P2W network of the modular system is a more powerful language model conditioned on given phoneme sequences. In this work, vanilla CTC and cross entropy with attention-based S2S are used as the criterion. \cite{liu2015} shows that the implicit language model of CTC can beat some weak explicit language models and the decoder of S2S has the same structure as the traditional LSTM language model. This indicates that the P2W network has the ability to model the extra text data. In general, the proposed P2W network could solve three issues:
\begin{itemize}
\item Complete phoneme alignment automatically;
\item Predict proper words in the dictionary with the given phoneme sequence;
\item Infer proper words with the given word sequence history.
\end{itemize}

\subsection{PSD Joint Training}
Finally, the P2W network will be fine-tuned on the acoustic data. In this phase, the A2P network would be fixed. In practice, the length of the acoustic feature (usually about one thousand) is much longer than the length of its corresponding phoneme sequence (usually less than one hundred). It is because the posterior sequence predicted by the A2P network and its corresponding phoneme sequence have different information rates. Therefore, it is not suitable to directly use the phoneme posteriors, which is the output of the A2P network as the input of the P2W network. Down-sampling is of help here. Phoneme synchronous decoding (PSD) \cite{chen2016} is a technique that is originally designed to speed up the decoding of CTC. It removes the \textbf{blank} frames in the phoneme posterior sequence, which can greatly reduce the information rate without performance loss. In this work, PSD is employed as a down-sampling layer between the A2P and P2W network. Given the A2P network output, i.e., the posterior sequence $\p_1,\ldots,\p_n$ where $\p_i(t)$ represents the posterior of phoneme $t$ at frame $i$, we have
\begin{align}
    \mathcal{X}=\{i|[\log\p_i(\text{\bf blank})-\max_{t\ne\text{\bf blank}}{\log\p_i(t)}]<\lambda\} \\
    \text{PSD}(\p_1,\ldots,\p_n) = \p_{k_1},\ldots,\p_{k_m} \quad k_i\in\mathcal{X}.
\end{align}
Here $\lambda$ is a pre-defined threshold. This means that PSD will remove the frames of A2P output that have high posterior on the {\bf blank} label.

The advantages of using PSD include:
\begin{itemize}
    \item Adapt different information rates among input acoustic features, intermediate phoneme sequence and predicted word sequence.
    \item Remove unnecessary \textbf{blank} information of the A2P network output.
    \item Speed up the training of P2W network.
\end{itemize}
With PSD, the whole system can be considered as \begin{align}P(\w|\x)&=\sum_{\p}P(\w|\p)P(\p|\x)\\
&\approx\text{P2W}(\text{PSD}(\text{A2P}(\x))).\end{align}

\subsection{Advantages of Modular E2E ASR System}
Compared with traditional HMM ASR systems, E2E systems could be easily deployed. The reason is that the E2E system only involves neural network calculations, which means only neural network parameters need to be stored. Besides, the neural network calculation can be accelerated by GPU or NPU, making it more suitable on some devices such as smartphones.
However, one big obstacle to the E2E ASR system is that its performance heavily relies on the amount of acoustic data. Compared with expensive acoustic data, text data is easier to collect. Phoneme or character based E2E systems try to improve prediction accuracy with the WFST encoded with language model trained on additional text data. This reduces the E2E property compared with HMM ASR systems. Here, the proposed modular design splits the whole system into two parts: the part that is only trained by acoustic data and the part that is trained by text data. In the training phase, the modular system is split into an acoustic model and a language model. During the decoding phase, the whole system performs as a unified A2W model. In general, the modular E2E ASR system is an A2W model that can be easily trained with extra text data. Moreover, its modular design enables system extension or adaption since we can fine-tune or re-train the A2P and P2W networks separately. In the next section, the OOV word extension is given as an example.

\section{Implementation Details}
\label{sec:imp}

This section exhibits the implementation details of the modular E2E ASR system. The OOV word extension is taken as an example to illustrate the extension capability of the proposed system.

\subsection{Phoneme Sequence Generation}
The large scale text data is used to train the P2W network. However, the P2W network needs the phoneme sequence posterior as the input, while the text data only contains word sequences. Therefore, the corresponding phoneme sequence of extra text data should be generated. In this work, a word to phoneme dictionary is utilized. For polyphone words, we randomly choose one pronunciation as the oracle one.

For each word $w$ in the word sequence, the dictionary is used to look up its corresponding phoneme sequence $\text{dict}(w)$, which are finally concatenated as the generated phoneme sequence. Formally, for a word sequence $\w=(w_1, \ldots, w_N)$ in extra text data, the generated phoneme sequence is \begin{equation}\p=\mathbf{1}(\text{concatenate}_{i=1}^T\text{dict}(w_i)).\end{equation} Here $\mathbf{1}$ means mapping the phoneme sequence to its one-hot distribution form. The data pair $(\p, \w)$ will be used to train the P2W network.

\subsection{Text Data Initialization}
\label{sec:tdi}
The generated phoneme sequence can be used to train the P2W network with large scale text data. However, during the decoding phase, the predicted phoneme posterior may be problematic since it is calculated by an A2P network rather than generated by a oracle word sequence. The mismatch between the phoneme sequence predicted by the A2P network and the oracle phoneme sequence generated by the word sequence will lead to performance degradation. In fact, the experiments show that training P2W network only with oracle phoneme sequences will lead to imprecise results.

To solve this problem, the oracle phoneme sequences generated by extra text data will be used to initialize the P2W network. After that, the P2W network is fine-tuned by predicted phoneme sequences with PSD.

Let $(\x_a,\w_a)$ denote the feature sequences and word sequences of acoustic data, and $\w_t$ denote the word sequences of text data. The whole training work-flow is shown below:

\begin{enumerate}
\item Use phoneme sequence generation method as described above to generate the corresponding oracle phoneme posterior sequences $\p_a^o$ and $\p_t^o$.
\item Use acoustic data $(\x_a, \p_a^o)$ to train the A2P network.
\item Use large scale text data $(\p_t^o, \w_t)$ to initialize P2W network.
\item Generate the predicted phoneme posterior sequence $\p_a=\text{PSD}(\text{A2P}(\x_a))$.
\item Fine-tune P2W network with data $(\p_a, \w_a)$.
\end{enumerate}

In this work, the A2P and P2W networks are not jointly trained. It is because that only acoustic data can be used to jointly train the A2P and P2W network. However, the A2P network is already trained by the acoustic data. Moreover, the acoustic data used to fine-tune the P2W network is down-sampled by PSD, which can greatly reduce the number of frames and accelerate the fine-tuning speed.

\subsection{OOV Words Extension}
How to deal with OOV words is a crucial issue for ASR systems, especially for A2W E2E ASR systems. In both CTC and S2S systems, to extend words, the words should be added into the vocabulary, and the dimension of the last softmax layer would be changed. Hence, to extend OOV words in a traditional A2W E2E ASR system, the whole neural network needs to be re-trained by acoustic data that contains OOV words. The time and resource consumption of the re-training procedure is huge. Another problem is that OOV words are usually rare words. The frequency of OOV words in the original acoustic data may be very low or even zero. This will lead to little performance improvement after re-training. Some approaches \cite{sennrich2016} have been proposed to solve the above problems. 
However, in this paper, we mainly focus on A2W models. \cite{sennrich2016} used sub-word as the output of the model, which is not an A2W E2E system. 

Here, the proposed modular E2E system can be used to solve the above OOV words extension problem. The most time-consuming part is the training of the acoustic model, i.e., the A2P network. However, the trained A2P network can be directly used to decode the phoneme for OOV words. Only the P2W network needs to be re-trained. It is noticeable that the P2W network can be trained by \textbf{TEXT} data. It indicates that to re-train the P2W network, the data which needs to be obtained is the text data containing the OOV words rather than acoustic data. Compared to acoustic data, text data is easier and cheaper to obtain.

More precisely, let $\D$ be the original acoustic data, i.e., $(\p_a, \w_a)$ in the above subsection and $\A$ indicate the extra text data containing the OOV words. Here, $\A$ contains the corresponding phoneme sequences by the same phoneme sequence generation method described above. Three ways are proposed to re-train the P2W network in a normally trained modular E2E ASR system.
\begin{enumerate}
\item \textit{Directly fine-tuning}: just use the extra text data $\A$ to fine-tune the P2W network.
\item \textit{Alternative training}: train the P2W network alternately between epochs by the original acoustic data $\D$ and extra text data $\A$.
\item \textit{Multi-modal}: this method \cite{renduchintala2018} is only proposed for S2S P2W network. An additional encoder is added to the P2W network. Samples from $\D$ are encoded by the original encoder and samples from $\A$ are encoded by the new encoder, while they are both decoded by the original decoder.
\end{enumerate}
More details of OOV words extension of modular E2E ASR system can be found in \cite{li2019oov}.

\section{Experiments}
\label{sec:exp}

\subsection{Experimental Setup}
The data corpus used for the experiments is SwitchBoard \cite{godfrey1992}. This corpus contains 300 hours of speech. The extracted acoustic feature is 36-dimensional fbank over 25ms time-window and 10ms frame shift. The neural networks are trained by Kaldi \cite{povey2011}, PyTorch \cite{paszke2017} and MXNet \cite{chen2015}.

The extra large scale text data is the transcription of Fisher corpus, which contains more than 2M sentences and 22M words. Both acoustic data and text data use the same vocabulary with size 30275. The evaluation sets are swbd and callhm from NIST eval2000 test set.

The baseline hybrid HMM model contains a 5-layer LSTM, while each layer contains 1024 memory cells and a 256 nodes projection layer. The last layer is softmax among 8k clustered tri-phone states.

The standard modular E2E ASR system consists of two networks. The A2P network contains 4-layer bidirectional LSTM, and each layer contains 1024 cells. The P2W network has two versions. The CTC version has a 3-layer bidirectional LSTM, and each BLSTM layer contains 1024 memory cells. The S2S version is composed of an encoder with 5-layer bidirectional LSTM and a decoder with a 5-layer unidirectional LSTM. Each layer of both encoder and decoder networks has 700 memory cells. Finally, the default PSD threshold is 8.

\subsection{Experimental Results of Modular E2E ASR System}
\subsubsection{Different P2W Training Procedure}
In the standard modular E2E training procedure, the P2W network needs to be trained twice. It is first initialized by large scale text data and then fine-tuned by the prediction of the A2P network. Some experiments are conducted to exhibit the necessity of extra text data and fine-tuning process. Table \ref{tab:p2w} shows the WER performance. It demonstrates that acoustic data fine-tuning of the P2W network is necessary for modular E2E systems. Even if the P2W network achieves a very low WER of 1.4/2.5 with oracle phoneme sequences, it fails to obtain the characteristics of the output of the A2P network and produces poor results. We also tried to remove the extra text data and trained the modular system with only the acoustic data. Then the whole system degraded to a normal A2W model. The performance is worse than the modular system trained with text data.

\begin{table}[thbp!]
    \centering
    \caption{WER performance comparison among different training procedures. TDI refers to text data initialization mentioned in section \ref{sec:tdi}. The numbers mean WER of test-set swbd and callhm, respectively.}
    \begin{tabular}{|c||c c|}
        \hline
        \textbf{Training Procedure} & \textbf{CTC} & \textbf{S2S}\\
        \hline
        No extra text data & 16.3/29.2 & 17.6/29.4\\
        No fine-tuning & 54.5/61.7 & 73.1/76.5 \\
        \hline
        TDI & 15.5/27.6 & 16.8/29.4 \\
        \hline
    \end{tabular}
    \label{tab:p2w}
\end{table}

\subsubsection{Different PSD Threshold}
The PSD threshold \cite{chan2016} controls the number of frames of training data for P2W network fine-tuning. With more frames, the information would be more complete while the training will be slower and vise versa. Here, some experiments are conducted to check the influence of the PSD threshold on the performance of modular E2E models. Table \ref{tab:psd} shows the results. The results show that a large PSD threshold can slightly improve performance. However, a too large or too small PSD threshold will make the P2W network more difficult to converge, especially for random initialized ones. It also demonstrates that extra text data can not only improve the performance but also enhances the convergence of the P2W network.

\begin{table}[thbp!]
    \centering
    \caption{WER performance comparison among different PSD thresholds. TDI means text data initialization, i.e. the using of extra text data.}
    \begin{tabular}{|c|c||c c c|}
        \hline
        \textbf{PSD}& \textbf{\# of Frames} & \textbf{TDI} & \textbf{CTC} & \textbf{S2S}\\
        \hline
        \multirow{2}*{3} & \multirow{2}*{12700679} & NO & 17.6/29.9 & 17.3/29.7 \\
        && YES & 16.6/30.9 & 17.0/29.7 \\
        \hline
        \multirow{2}*{8} & \multirow{2}*{15997606} & NO & 16.3/29.2 & 17.6/29.4 \\
        && YES & 15.5/27.6 & 16.8/29.4 \\
        \hline
        \multirow{2}*{15} & \multirow{2}*{22270884} & NO &  19.6/29.3 & 17.6/30.1 \\
       &&  YES & 15.4/27.1 & 16.7/29.6 \\
        \hline        
    \end{tabular}
    \label{tab:psd}
\end{table}

\subsubsection{Different Acoustic Models}
Other than the baseline BLSTM acoustic model, two weak acoustic models are trained to examine the effect of different acoustic models. The weak LSTM acoustic model contains a 5-layer LSTM, each LSTM layer has 1024 memory cells and 256 projection nodes \cite{sak2014}. The weak FSMN model contains 8-layer FSMN \cite{zhang2015} and 2-layer DNN. Each FSMN layer contains 1024 units and 256 projection nodes \cite{sak2014}, while each DNN layer contains 1024 units. Each acoustic model is trained by a CTC criterion based on phoneme. The phoneme CTC system is directly decoded by an extra WFST encoded with a language model. The word CTC system has the same structure as the phoneme CTC system except for the last softmax layer. The modular systems use the trained acoustic model as the A2P network. Table \ref{tab:a2p} shows the WER results. It is clear that modular systems can be improved by using better A2P models. This means all the optimization methods used on normal E2E acoustic models could be used to improve the performance of the A2P network, which may lead to performance improvement of the proposed modular system.

\begin{table}[thbp!]
    \centering
    \caption{WER performance comparison between different acoustic models. Text indicates does the model use extra text data. A2W indicates is this model an A2W model during the decoding phase.}
    \begin{tabular}{|c|c||c|c||c|}
        \hline
        \multicolumn{2}{|c||}{\textbf{Model}} & \textbf{A2W} & \textbf{Text} & \textbf{WER} \\
        \hline
        \multirow{4}*{LSTM} & Phoneme CTC & \xmark & \cmark & 19.4/33.5 \\
        & Word CTC &  \cmark & \xmark & 29.6/31.7 \\        
        & Modular CTC & \cmark & \cmark & 19.4/30.7 \\
        & Modular S2S & \cmark & \cmark & 21.3/35.1 \\
        \hline
        \multirow{4}*{FSMN} & Phoneme CTC & \xmark & \cmark & 14.1/26.3 \\
        & Word CTC &  \cmark & \xmark & 23.1/34.9 \\        
        & Modular CTC & \cmark & \cmark & 17.5/28.7 \\
        & Modular S2S & \cmark & \cmark & 19.6/31.5 \\
        \hline
        \multirow{4}*{BLSTM} & Phoneme CTC & \xmark & \cmark & 12.8/24.0 \\
        & Word CTC &  \cmark & \xmark & 21.1/31.4 \\        
        & Modular CTC & \cmark & \cmark & 15.5/27.6 \\
        & Modular S2S & \cmark & \cmark & 16.8/29.4 \\
        \hline        
    \end{tabular}
    \label{tab:a2p}
\end{table}

We could also observe that the performance of the modular S2S is slightly worse than the modular CTC model. The reason is that the output of A2P might contain errors in the predicted word sequence. Since CTC assumes that each predicted word is independent of the other words, the errors would only affect a small area. However, in S2S models, each word prediction depends on its predecessors. The errors will accumulate to degrade the final performance. In fact, if we use the oracle phoneme sequence as model input, S2S will perform better than CTC. It is also found that the gap between modular CTC and modular S2S system is smaller with a better acoustic model from table \ref{tab:a2p}.

\subsubsection{Comparison Among Different Baselines}
We have compared the WER performance among different ASR baselines, including the DNN-HMM hybrid system, the two types of traditional E2E systems, and the proposed modular system. Table \ref{tab:e2e} shows the comparison results. It can be observed that the E2E systems with small units output like character or phoneme perform better than normal A2W models. It also can be observed that this type of E2E systems can be easily combined with a language model to improve their performance. For A2W models, the proposed modular systems perform better than normal word CTC or S2S models. It is believed that the performance improvements come from the use of extra text data. Overall, the proposed modular E2E model can get better performance by the extra text data compared with traditional A2W models, i.e., the A2W models that are directly trained by CTC or S2S with word-level output units. Compared with other types of E2E systems that are based on characters or phonemes, the proposed modular systems achieve slightly worse performance while holding the A2W E2E property, i.e., the whole system performs as a single A2W neural network during decoding phase.

\begin{table}[thbp!]
    \centering
    \caption{WER performance comparison among different baselines. Text indicates does the model use extra text data. A2W indicates is this model an A2W model during the decoding phase. The models with an asterisk are trained by us.}
    \begin{tabular}{|c||c|c||c|}
        \hline
        \textbf{Model} & \textbf{A2W} & \textbf{Text} & \textbf{WER} \\
        \hline
        Hybrid HMM*  & \xmark & \cmark & 14.9/27.6 \\
        \hline
        Phoneme CTC\cite{toshniwal2017} & \xmark & \cmark & 24.6/41.3 \\
        Character CTC\cite{audhkhasi2018} & \xmark & \xmark & 18.9/30.9 \\
        Character CTC\cite{zweig2017} & \xmark & \cmark & 14.0/25.3 \\
        Phoneme CTC* & \xmark & \cmark & 12.8/24.0 \\
        \hline
        Phoneme S2S\cite{toshniwal2017} & \xmark & \cmark & 23.1/40.8 \\
        Character S2S\cite{lu2016} & \xmark & \xmark & 26.8/48.2 \\
        Character S2S\cite{palaskar2018} & \xmark & \cmark & 15.6/31.0 \\
        Character S2S* & \xmark & \cmark & 16.7/30.3 \\
        \hline
        Word CTC\cite{audhkhasi2018} & \cmark & \xmark & 17.4/26.9 \\
        Word S2S\cite{palaskar2018} & \cmark & \xmark & 22.4/36.2 \\
        Word S2S\cite{palaskar2018} & \cmark & \cmark & 22.1/36.3 \\
        \hline
        Modular CTC* & \cmark & \cmark & 15.5/27.6 \\
        Modular S2S* & \cmark & \cmark & 16.8/29.4 \\
        \hline
    \end{tabular}
    \label{tab:e2e}
\end{table}

\cite{audhkhasi2018} obtains better results by using multiple optimization methods. These methods, including speed perturb, i-vector adaption, phoneme network initialization, and CTC-S2S joint training, are not used in this work. It is believed that the proposed modular E2E ASR system could achieve better results with these optimization algorithms.

\subsection{Experimental Results of OOV Words Extension}
This subsection demonstrates the modular design can do extension or adaption easily. Here OOV word extension is used as an example.

\subsubsection{Extra Test Set}
The data corpus used for the OOV word extension is almost the same as normal modular E2E experiments. The only difference is the evaluation set. Two data corpora are used. One is the in-domain eval2000 test set which is the combination of swbd and callhm, and the other is the cross-domain dev93 test set from WSJ corpus.

To investigate the performance gain of the OOV words extension, the full vocabulary $\V_f$ has been cut to small vocabulary $\V_s$ where $\V_s$ only contains words that occur more than 10 times in the training set. The baseline modular E2E models are trained with the small vocabulary $\V_s$. And the OOV words extension models are trained with vocabulary $\V_{ev}$ and $\V_{dev}$, which are the union of $\V_s$ and the corresponding vocabulary of each test set. Table \ref{tab:vocab} shows the OOV rate of each vocabulary. The extra text data is extracted from Fisher corpus. Here, not all the sentences in Fisher are used. Only the sentences containing OOV words are used. 

\begin{table}[thbp!]
    \centering
    \caption{The size and OOV rate of each vocabulary.}
    \begin{tabular}{|c|c||c|c|c|}
        \hline
        \multirow{3}{*}{\textbf{Vocabulary}} & \multirow{3}{*}{\textbf{Size}} & \multicolumn{3}{c|}{\textbf{OOV Rate}} \\
        \cline{3-5}
        & & \textbf{Training} & \textbf{Eval2000} & \textbf{Dev93} \\
        & & \textbf{Set} & \textbf{Test Set} & \textbf{Test Set} \\
        \hline
        $\V_f$ & 30275 & 0 & 1.47 & 6.4 \\
        \hline
        $\V_s$ & 6805 & \multirow{3}{*}{2.04} & 3.33 & 15.2 \\
        \cline{1-2} \cline{4-5}
        $\V_{ev}$ & 7649 & & 0.27 & - \\
        \cline{1-2} \cline{4-5}
        $\V_{dev}$ & 7627 & & - & 1.2 \\
        \hline
    \end{tabular}
    \label{tab:vocab}
\end{table}

The test set is split into two subsets, which are in vocabulary sentences (IVS) and out of vocabulary sentences (OOVS). If every word in a sentence appears in vocabulary $\V_s$ , then this sentence belongs to IVS and otherwise OOVS.

\subsubsection{Experimental Results of In-domain Test Set}

Table \ref{tab:eval} shows the WER performance on in-domain eval2000 test set with OOV words extension. It shows that for the in-domain test set, the baseline system with full vocabulary performs better than the system with a small vocabulary. It is also observed that for the in-domain test set, the OOV words do not cause much performance degradation. The WER gap between IVS and OOVS is about 20\% relatively. And since the OOV rate is 3.33\% for the baseline small vocabulary models, the poor WER on OOVS has a small impact on the total WER.

\begin{table}[thbp!]
    \centering
    \caption{WER performance comparison on in-domain eval2000 with OOV words extension.}
    \begin{tabular}{|l|c||c c c|}
        \hline
        \multicolumn{1}{|c|}{\multirow{2}{*}{\textbf{Model}}} & \textbf{Vocabulary} & \multicolumn{3}{c|}{\textbf{WER}} \\
        \cline{3-5}
        & \textbf{Size} & \textbf{All} & \textbf{IVS} & \textbf{OOVS} \\
        \hline
        \multirow{2}{*}{Modular CTC} & 30275 & 22.8 & 21.0 & 26.0 \\
        & 6805 & 26.0 & 24.4 & 29.0 \\
        \hline
        \quad + directly fine-tuning & \multirow{2}{*}{7649} & 24.6 & 23.0 & 27.5 \\
        \quad + alternative training & & 23.5 & 22.5 & 25.4 \\
        \hline
        \multirow{2}{*}{Modular S2S} & 30275 & 23.5 & 21.1 & 28.0 \\
        & 6805 & 24.5 & 22.1 & 29.0 \\
        \hline
        \quad + directly fine-tuning & \multirow{3}{*}{7649} & 26.8 & 24.0 & 32.1 \\
        \quad + alternative training & & 24.2 & 22.0 & 28.3 \\
        \quad + multi-modal & & 24.3 & 21.7 & 29.2 \\
        \hline        
    \end{tabular}
    \label{tab:eval}
\end{table}

For CTC models, directly fine-tuning and alternative training all outperform the small baseline models. The alternative training model even beats the large baseline model on OOVS. For S2S models, the gap between small and large baseline models is smaller than CTC, this indicates S2S models are influenced less by OOV words compared with CTC models. Directly fine-tuning only uses the text data with OOV words may increase the risk of over-fitting for S2S models. However, alternative training still performs slightly better than the small baseline system. Multi-modal is less effective compared with alternative training, especially on OOVS.

\subsubsection{Experimental Results of Cross-domain Test Set}

Table \ref{tab:dev} gives the results of the cross-domain dev93 test set from WSJ. Since no acoustic data from WSJ are used to train the models, the WER is much higher than other reported results. It can be observed that the WER of OOVS is much higher than IVS. Given that the OOV rate of small baseline systems is 15.2\% and 6.4\% for big baseline systems, the overall WER is influenced a lot by the poor WER of OOVS. It is also observed that the baseline S2S performs poorly due to the predicted word dependence during decoding.

\begin{table}[thbp!]
    \centering
    \caption{WER performance comparison on cross-domain dev93 with OOV words extension.}
    \begin{tabular}{|l|c||c c c|}
        \hline
        \multicolumn{1}{|c|}{\multirow{2}{*}{\textbf{Model}}} & \textbf{Vocabulary} & \multicolumn{3}{c|}{\textbf{WER}} \\
        \cline{3-5}
        & \textbf{Size} & \textbf{All} & \textbf{IVS} & \textbf{OOVS} \\
        \hline
        \multirow{2}{*}{Modular CTC} & 30275 & 39.1 & 26.2 & 40.8 \\
        & 6805 & 36.4 & 18.4 & 38.7 \\
        \hline
        \quad + directly fine-tuning & \multirow{2}{*}{7627} & 36.9 & 17.6 & 39.4 \\
        \quad + alternative training & & 30.3 & 17.8 & 31.9 \\
        \hline
        \multirow{2}{*}{Modular S2S} & 30275 & 43.8 & 22.6 & 46.5 \\
        & 6805 & 41.3 & 20.4 & 44.0 \\
        \hline
        \quad + directly fine-tuning & \multirow{3}{*}{7627} & 39.1 & 20.5 & 42.2 \\ 
        \quad + alternative training & & 35.6 & 18.9 & 37.8 \\ 
        \quad + multi-modal & & 40.7 & 18.5 & 43.6 \\ 
        \hline     
    \end{tabular}
    \label{tab:dev}
\end{table}

For cross-domain experiments, the usage of full vocabulary does not work well. For CTC models, alternative training improves the performance of OOVS compared with both baseline systems. It even gets better results on IVS. For modular S2S models, directly fine-tuning and alternative training outperform the baseline systems on both IVS and OOVS. Lastly, it can be seen that multi-modal is still ineffective on OOVS.

Overall, using a small vocabulary will degrade the performance. Extending the vocabulary to full size is beneficial for in-domain tasks, but not performs well for cross-domain tasks. However, the proposed modular E2E ASR system could use extra text data to extend the OOV words to improve the performance. Regarding these three fine-tuning methods, only multi-modal is considered to be ineffective on OOVS. Directly fine-tuning in useful on OOVS in almost every case. And alternative training is the best choice. However, it needs more time to be trained and is harder to converge.

\section{Conclusion}
\label{sec:conc}
This paper proposes a modular training strategy for E2E ASR. In particular, the proposed method splits the E2E systems into two parts: A2P and P2W networks. The P2W networks can be trained with large scale text data, which can improve the WER performance. During the decoding phase, the two networks are combined together and act as a single A2W network that holds the E2E property. Experiments on 300 hours SwitchBoard corpus show that this novel approach outperforms the naive A2W models and reaches the level of state-of-the-art A2W \cite{audhkhasi2018} models with the same training procedure. Besides, the modular design enables the efficient revision of the whole system. The OOV words extension experiment provides an example. The future work includes:
\begin{itemize}
\item Use other optimization methods including speed perturb \cite{ko2015}, speaker adaptation \cite{tan2015} and GloVe initialization \cite{pennington2014} to improve the A2P network;
\item Train P2W network with other E2E criteria such as CTC-S2S multitask \cite{kim2017} method.
\item Try other cross-domain experiments by using the same method in the OOV words extension.
\end{itemize}


\bibliographystyle{IEEEtran}
\bibliography{refs}

\begin{thebibliography}{10}
\providecommand{\url}[1]{#1}
\csname url@samestyle\endcsname
\providecommand{\newblock}{\relax}
\providecommand{\bibinfo}[2]{#2}
\providecommand{\BIBentrySTDinterwordspacing}{\spaceskip=0pt\relax}
\providecommand{\BIBentryALTinterwordstretchfactor}{4}
\providecommand{\BIBentryALTinterwordspacing}{\spaceskip=\fontdimen2\font plus
\BIBentryALTinterwordstretchfactor\fontdimen3\font minus
  \fontdimen4\font\relax}
\providecommand{\BIBforeignlanguage}[2]{{%
\expandafter\ifx\csname l@#1\endcsname\relax
\typeout{** WARNING: IEEEtran.bst: No hyphenation pattern has been}%
\typeout{** loaded for the language `#1'. Using the pattern for}%
\typeout{** the default language instead.}%
\else
\language=\csname l@#1\endcsname
\fi
#2}}
\providecommand{\BIBdecl}{\relax}
\BIBdecl

\bibitem{morgan1990}
N.~Morgan and H.~Bourlard, ``Continuous speech recognition using multilayer
  perceptrons with hidden markov models,'' in \emph{Proceedings of
  International Conference on Acoustics, Speech and Signal Processing
  (ICASSP)}, 1990, pp. 413--416.

\bibitem{bourlard1992}
H.~Bourlard and C.~J. Wellekens, ``Links between markov models and multilayer
  perceptrons,'' \emph{IEEE Transactions on Pattern Analysis and Machine
  Intelligence}, vol.~12, no.~12, pp. 1167--1178, 1990.

\bibitem{morgan1995}
N.~Morgan and H.~Bourlard, ``Neural networks for statistical recognition of
  continuous speech,'' \emph{Proceedings of the IEEE}, vol.~83, no.~5, pp.
  742--772, 1995.

\bibitem{mohria2002}
M.~Mohria, F.~Pereirab, and M.~Rileya, ``Weighted finite-state transducers in
  speech recognition,'' \emph{Computer Speech \& Language}, vol.~16, no.~1, pp.
  69--88, 2002.

\bibitem{sak2014}
H.~Sak, A.~Senior, and F.~Beaufays, ``Long short-term memory recurrent neural
  network architectures for large scale acoustic modeling,'' in
  \emph{Proceedings of Annual Conference of the International Speech
  Communication Association (INTERSPEECH)}, 2014, pp. 338--342.

\bibitem{qian2016}
Y.~M. Qian, M.~X. Bi, T.~Tan, and K.~Yu, ``Very deep convolutional neural
  networks for noise robust speech recognition,'' \emph{IEEE/ACM Transactions
  on Audio, Speech, and Language Processing}, vol.~24, no.~12, pp. 2263--2276,
  2016.

\bibitem{graves2006}
A.~Graves, S.~Fern{\'a}ndez, F.~Gomez, and J.~Schmidhuber, ``Connectionist
  temporal classification: Labelling unsegmented sequence data with recurrent
  neural networks,'' in \emph{Proceedings of International Conference on
  Machine Learning (ICML)}, 2006, pp. 369--376.

\bibitem{graves2014}
A.~Graves and N.~Jaitly, ``Towards end-to-end speech recognition with recurrent
  neural networks,'' in \emph{Proceedings of International Conference on
  Machine Learning (ICML)}, 2014, pp. 1764--1772.

\bibitem{liu2015}
Q.~Liu, L.~J. Wang, and Q.~Huo, ``A study on effects of implicit and explicit
  language model information for {DBLSTM-CTC} based handwriting recognition,''
  in \emph{Proceedings of International Conference on Document Analysis and
  Recognition (ICDAR)}, 2015, pp. 461--465.

\bibitem{sutskever2014}
I.~Sutskever, O.~Vinyals, and Q.~V. Le, ``Sequence to sequence learning with
  neural networks,'' in \emph{Proceedings of Neural Information Processing
  Systems Conference (NIPS)}, 2014, pp. 3104--3112.

\bibitem{cui2017}
Y.~M. Cui, Z.~P. Chen, S.~Wei, S.~J. Wang, T.~Liu, and G.~Hu,
  ``Attention-over-attention neural networks for reading comprehension,'' in
  \emph{Proceedings of Annual Meeting of the Association for Computational
  Linguistics (ACL)}, 2017, pp. 593--602.

\bibitem{chan2016}
W.~Chan, N.~Jaitly, Q.~V. Le, and O.~Vinyals, ``Listen, attend and spell: A
  neural network for large vocabulary conversational speech recognition,'' in
  \emph{Proceedings of International Conference on Acoustics, Speech and Signal
  Processing (ICASSP)}, 2016, pp. 4960--4964.

\bibitem{bahdanau2016}
D.~Bahdanau, J.~Chorowski, D.~Serdyuk, P.~Brakel, and Y.~Bengio, ``End-to-end
  attention-based large vocabulary speech recognition,'' in \emph{Proceedings
  of International Conference on Acoustics, Speech and Signal Processing
  (ICASSP)}, 2016, pp. 4945--4949.

\bibitem{graves2012}
A.~Graves, \emph{Sequence Transduction with Recurrent Neural Networks},
  arXiv:1211.3711, 2012.

\bibitem{graves2013}
A.~Graves, A.~r.~Mohamed, and G.~Hinton, ``Speech recognition with deep
  recurrent neural networks,'' in \emph{Proceedings of International Conference
  on Acoustics, Speech and Signal Processing (ICASSP)}, 2013, pp. 6645--6649.

\bibitem{battenberg2017}
E.~Battenberg, J.~T. Chen, R.~Child, A.~Coates, Y.~Gaur, Y.~Li, H.~R. Liu,
  S.~Satheesh, A.~Sriram, and Z.~Y. Zhu, ``Exploring neural transducers for
  end-to-end speech recognition,'' in \emph{Proceedings of Automatic Speech
  Recognition and Understanding Workshop (ASRU)}, 2017, pp. 206--213.

\bibitem{miao2015}
Y.~J. Miao, M.~Gowayyed, and F.~Metze, ``{EESEN}: End-to-end speech recognition
  using deep {RNN} models and {WFST}-based decoding,'' in \emph{Proceedings of
  Automatic Speech Recognition and Understanding Workshop (ASRU)}, 2015, pp.
  167--174.

\bibitem{chen2016}
Z.~H. Chen, W.~Deng, T.~Xu, and K.~Yu, ``Phone synchronous decoding with {CTC}
  lattice,'' in \emph{Proceedings of Annual Conference of the International
  Speech Communication Association (INTERSPEECH)}, 2016, pp. 1923--1927.

\bibitem{hori2018}
T.~Hori, J.~Cho, and S.~Watanabe, \emph{End-to-end Speech Recognition with
  Word-based {RNN} Language Models}, arXiv:1808.02608, 2018.

\bibitem{hayashi2018}
T.~Hayashi, S.~Watanabe, Y.~Zhang, T.~Toda, T.~Hori, R.~Astudillo, and
  K.~Takeda, \emph{Back-Translation-Style Data Augmentation for End-to-End
  {ASR}}, arXiv:1807.10893, 2018.

\bibitem{renduchintala2018}
A.~Renduchintala, S.~Y. Ding, M.~Wiesner, and S.~Watanabe, ``Multi-modal data
  augmentation for end-to-end {ASR},'' in \emph{Proceedings of Annual
  Conference of the International Speech Communication Association
  (INTERSPEECH)}, 2018, pp. 2394--2398.

\bibitem{sennrich2016}
R.~Sennrich, B.~Haddow, and A.~Birch, ``Neural machine translation of rare
  words with subword units,'' in \emph{Proceedings of Annual Meeting of the
  Association for Computational Linguistics (ACL)}, 2016, pp. 1715--1725.

\bibitem{chen2018}
Z.~H. Chen, Q.~Liu, H.~Li, and K.~Yu, ``On modular training of neural
  acoustics-to-word model for {LVCSR},'' in \emph{Proceedings of International
  Conference on Acoustics, Speech and Signal Processing (ICASSP)}, 2018, pp.
  4754--4758.

\bibitem{baum1967}
L.~Baum and J.~Eagon, ``An inequality with applications to statistical
  estimation for probabilistic functions of markov processes and to a model for
  ecology,'' \emph{Bulletin of the American Mathematical Society}, vol.~73,
  no.~3, pp. 360--363, 1967.

\bibitem{baum1968}
L.~Baum and G.~Sell, ``Growth transformations for functions on manifolds,''
  \emph{Pacific Journal of Mathematics}, vol.~27, no.~2, pp. 211--227, 1968.

\bibitem{bahdanau2015}
D.~Bahdanau, K.~Cho, and Y.~Bengio, ``Neural machine translation by jointly
  learning to align and translate,'' in \emph{Proceedings of International
  Conference on Learning Representations (ICLR)}, 2015.

\bibitem{variani2017}
E.~Variani, T.~Bagby, E.~McDermott, and M.~Bacchiani, ``End-to-end training of
  acoustic models for large vocabulary continuous speech recognition with
  tensorflow,'' in \emph{Proceedings of Annual Conference of the International
  Speech Communication Association (INTERSPEECH)}, 2017, pp. 1641--1645.

\bibitem{lu2016}
L.~Lu, X.~Zhang, and S.~Renais, ``On training the recurrent neural network
  encoder-decoder for large vocabulary end-to-end speech recognition,'' in
  \emph{Proceedings of International Conference on Acoustics, Speech and Signal
  Processing (ICASSP)}, 2016, pp. 5060--5064.

\bibitem{miao2016}
Y.~Miao, M.~Gowayyed, X.~Na, T.~Ko, F.~Metze, and A.~Waibel, ``An empirical
  exploration of {CTC} acoustic models,'' in \emph{Proceedings of International
  Conference on Acoustics, Speech and Signal Processing (ICASSP)}, 2016, pp.
  2623--2627.

\bibitem{watanabe2018}
S.~Watanabe, T.~Hori, S.~Karita, T.~Hayashi, J.~Nishitoba, Y.~Unno, N.~Soplin,
  J.~Heymann, M.~Wiesner, N.~Chen, A.~Renduchintala, and T.~Ochiai, ``{ESPnet}:
  End-to-end speech processing toolkit,'' in \emph{Proceedings of Annual
  Conference of the International Speech Communication Association
  (INTERSPEECH)}, 2018, pp. 2207--2211.

\bibitem{rao2017}
K.~Rao, H.~Sak, and R.~Prabhavalkar, ``Exploring architectures, data and units
  for streaming end-to-end speech recognition with rnn-transducer,'' in
  \emph{Proceedings of Automatic Speech Recognition and Understanding Workshop
  (ASRU)}, 2017, pp. 193--199.

\bibitem{toshniwal2017}
S.~Toshniwal, H.~Tang, L.~Lu, and K.~Livescu, ``Multitask learning with
  low-level auxiliary tasks for encoder-decoder based speech recognition,'' in
  \emph{Proceedings of Annual Conference of the International Speech
  Communication Association (INTERSPEECH)}, 2017, pp. 3532--3536.

\bibitem{audhkhasi2017}
K.~Audhkhasi, B.~Ramabhadran, G.~Saon, M.~Picheny, and D.~Nahamoo, ``Direct
  acoustics-to-word models for english conversational speech recognition,'' in
  \emph{Proceedings of Annual Conference of the International Speech
  Communication Association (INTERSPEECH)}, 2017, pp. 959--963.

\bibitem{audhkhasi2018}
K.~Audhkhasi, B.~Kingsbury, B.~Ramabhadran, G.~Saon, and M.~Picheny, ``Building
  competitive direct acoustics-to-word models for english conversational speech
  recognition,'' in \emph{Proceedings of International Conference on Acoustics,
  Speech and Signal Processing (ICASSP)}, 2018, pp. 4759--4763.

\bibitem{yu2018}
C.~Z. Yu, C.~L. Zhang, C.~Weng, J.~Cui, and D.~Yu, ``A multistage training
  framework for acoustic-to-word model,'' in \emph{Proceedings of Annual
  Conference of the International Speech Communication Association
  (INTERSPEECH)}, 2018, pp. 786--790.

\bibitem{he2019}
Y.~Z. He, T.~Sainath, R.~Prabhavalkar, I.~McGraw, R.~Alvarez, D.~Zhao,
  D.~Rybach, A.~Kannan, Y.~H. Wu, R.~M. Pang, Q.~Liang, D.~Bhatia,
  Y.~Shangguan, B.~Li, G.~Pundak, K.~C. Sim, T.~Bagby, S.~Y. Chang, K.~Rao, and
  A.~Gruenstein, ``Streaming end-to-end speech recognition for mobile
  devices,'' in \emph{Proceedings of International Conference on Acoustics,
  Speech and Signal Processing (ICASSP)}, 2019, pp. 6381--6385.

\bibitem{li2019}
J.~Y. Li, R.~Zhao, H.~Hu, and Y.~F. Gong, ``Improving {RNN} transducer modeling
  for end-to-end speech recognition,'' in \emph{Proceedings of Automatic Speech
  Recognition and Understanding Workshop (ASRU)}, 2019, pp. 114--121.

\bibitem{sriram2017}
A.~Sriram, H.~Jun, S.~Satheesh, and A.~Coates, \emph{Cold fusion: Training
  seq2seq models together with language models}, arXiv:1708.06426, 2017.

\bibitem{li2019oov}
H.~Li, Z.~H. Chen, Q.~Liu, Y.~M. Qian, and K.~Yu, ``{OOV} words extension for
  modular neural acoustics-to-word model,'' in \emph{Proceedings of National
  Conference on Man-Machine Speech Communication (NCMMSC)}, 2019.

\bibitem{godfrey1992}
J.~J. Godfrey, E.~C. Holliman, and J.~McDaniel, ``{SWITCHBOARD}: Telephone
  speech corpus for research and development,'' in \emph{Proceedings of
  International Conference on Acoustics, Speech and Signal Processing
  (ICASSP)}, 1992, pp. 517--520.

\bibitem{povey2011}
D.~Povey, A.~Ghoshal, G.~Boulianne, L.~Burget, O.~Glembek, N.~Goel,
  M.~Hannemann, P.~Motlicek, Y.~M. Qian, P.~Schwarz, J.~Silovsky, G.~Stemmer,
  and K.~Vesely, ``The kaldi speech recognition toolkit,'' in \emph{Proceedings
  of Automatic Speech Recognition and Understanding Workshop (ASRU)}, 2011.

\bibitem{paszke2017}
A.~Paszke, S.~Gross, S.~Chintala, G.~Chanan, E.~Yang, Z.~DeVito, Z.~M. Lin,
  A.~Desmaison, L.~Antiga, and A.~Lerer, \emph{Automatic Differentiation in
  PyTorch}, 2017.

\bibitem{chen2015}
T.~Q. Chen, M.~Li, Y.~T. Li, M.~Lin, N.~Y. Wang, M.~J. Wang, T.~J. Xiao, B.~Xu,
  C.~Y. Zhang, and Z.~Zhang, \emph{MXNet: A Flexible and Efficient Machine
  Learning Library for Heterogeneous Distributed Systems}, arXiv:1512.01274,
  2015.

\bibitem{zhang2015}
S.~L. Zhang, C.~Liu, H.~Jiang, S.~Wei, L.~R. Dai, and Y.~Hu, \emph{Feedforward
  Sequential Memory Networks: A New Structure to Learn Long-term Dependency},
  arXiv:1512.08301, 2015.

\bibitem{zweig2017}
G.~Zweig, C.~Yu, J.~Droppo, and A.~Stolcke, ``“advances in all-neural speech
  recognition,'' in \emph{Proceedings of International Conference on Acoustics,
  Speech and Signal Processing (ICASSP)}, 2017, pp. 4805--4809.

\bibitem{palaskar2018}
S.~Palaskar and F.~Metze, ``Acoustic-to-word recognition with
  sequence-to-sequence models,'' in \emph{Proceedings of IEEE Spoken Language
  Technology Workshop (SLT)}, 2018, pp. 397--404.

\bibitem{ko2015}
T.~Ko, V.~Peddinti, D.~Povey, and S.~Khudanpur, ``Audio augmentation for speech
  recognition,'' in \emph{Proceedings of Annual Conference of the International
  Speech Communication Association (INTERSPEECH)}, 2015, pp. 3586--3589.

\bibitem{tan2015}
T.~Tan, Y.~M. Qian, M.~F. Yin, Y.~M. Zhuang, and K.~Yu, ``Cluster adaptive
  training for deep neural network,'' in \emph{Proceedings of International
  Conference on Acoustics, Speech and Signal Processing (ICASSP)}, 2015, pp.
  4325--4329.

\bibitem{pennington2014}
J.~Pennington, R.~Socher, and C.~D. Manning, ``Glove: Global vectors for word
  representation,'' in \emph{Proceedings of Conference on Empirical Methods in
  Natural Language Processing (EMNLP)}, 2014, pp. 1532--1543.

\bibitem{kim2017}
S.~Kim, T.~Hori, and S.~Watanabe, ``Joint {CTC}-attention based end-to-end
  speech recognition using multi-task learning,'' in \emph{Proceedings of
  International Conference on Acoustics, Speech and Signal Processing
  (ICASSP)}, 2017, pp. 4835--4839.

\end{thebibliography}

\end{document}